\documentclass{ws-p8-50x6-00}

\begin{document}

\title{String Representation of Gauge Theories}

\author{Dmitri Antonov}

\address{INFN-Sezione di Pisa, Universit\'a degli studi di Pisa,
Dipartimento di Fisica, Via Buonarroti, 2 - Ed. B - 56127 Pisa, 
Italy\\
and\\
Institute of Theoretical and Experimental Physics, 
B. Cheremushkinskaya 25, RU-117 218 Moscow, Russia\\
E-mail: antonov@difi.unipi.it}


\maketitle

\abstracts{In this talk, various approaches to the problem 
of evaluation of the field strength correlators in the 
$SU(3)$-gluodynamics, which play the 
major r\^ole in the  
Stochastic Vacuum Model, are reviewed. This is done in the framework
of the effective Abelian-projected theories under the various 
assumptions implied on the properties of the ensemble of 
Abelian-projected monopoles. In particular, within the 
assumption on the condensation of the monopole Cooper pairs, 
the main method of investigation
is the string representation of field strength correlators.
The calculation of the bilocal field strength correlator
in the 3D effective theory, where Abelian-projected
monopoles are assumed to form a gas, based on the string 
representation for the Wilson loop in this theory, 
is also presented.}

\section{Introduction. 
The problem of string representation of gauge theories}
The problem of string representation of gauge theories is 
unambiguously related to the problem of confinement in these theories.
Its essence is the quest of a string theory, which is mostly adequate
for the description of strings between color objects, 
which appear in the confining phase of QCD. (Other, Abelian-type,  
gauge theories possessing the confining phase then serve  
for probing various approaches to the construction of 
the string representation of QCD.) 
Quantitatively, the QCD string can be seen by virtue 
of the Wilson's picture of confinement~\cite{1}. It states that
the criterion of confinement in QCD is the area law behavior
of the Wilson loop 

\begin{equation}
\label{wilson}
\left<W(C)\right>
\equiv\frac{1}{N_c}\left<{\rm tr}{\,}{\cal P}{\,}
\exp\left(ig\oint\limits_{C}^{}A_\mu^a T^adx_\mu\right)\right>
\stackrel{{C\to\infty}}{\longrightarrow}
{\rm e}^{-\sigma\left|\Sigma_{\rm min}(C)\right|}.
\end{equation}
Here, $\sigma$ 
is the so-called string tension, {\it i.e.} the energy density of 
the QCD string. The latter one is nothing else, but a tube 
formed by the lines of chromoelectric flux, which appears 
between two color objects propagating along the contour $C$.
When these objects try to move apart from each other, the QCD 
string stretches and prevents that, thus ensuring their confinement.
According to Eq.~(\ref{wilson}), during its propagation, such a string 
sweeps out the surface of the minimal area for 
a given contour $C$, $\Sigma_{\rm min}(C)$. Due to the dimensional
reasons, 

$$\sigma\propto\Lambda_{\rm QCD}^2=\frac{1}{a^2}\exp\left[-
\frac{16\pi^2}{\left(\frac{11}{3}N_c-\frac23N_f\right)g^2\left(a^{-2}
\right)}\right]$$
with $a\to 0$ standing for the distance UV cutoff ({\it e.g.} the lattice 
spacing). Clearly, all the coefficients in the expansion of 
$\sigma$ in powers of $g^2$ vanish, which means that the QCD string
is an essentially nonperturbative object.

Owing to this observation, it is nowadays 
commonly argued that the area law is well saturated 
by the strong background fields in QCD. Around those, there 
additionaly exist perturbative fluctuations of the 
QCD vacuum~\cite{2}, which  
excite the string. This means that these fluctuations enable the string  
to sweep out with a 
nonvanishing probability not only $\Sigma_{\rm min}(C)$, but
also an arbitrary surface $\Sigma(C)$, bounded by $C$. Therefore, 
the final aim in constructing the string representation 
of QCD is a derivation 
of the formula $\left<W(C)\right>=\sum\limits_{\Sigma(C)}^{}
{\rm e}^{-{\cal S}[\Sigma(C)]}$. Here, $\sum\limits_{\Sigma(C)}^{}$ and 
${\cal S}[\Sigma(C)]$ stand for a certain sum over string world sheets
and a string effective action, both of which are yet unknown in QCD.
Clearly, this formula is just a 2D analogue of the 
well known representation for the propagator of a point-like particle,
which is subject to external forces and/or propagates in external fields. 
In particular, the r\^ole of the classical
trajectory of such a particle is played within this analogy by 
$\Sigma_{\rm min}(C)$. However, it unfortunately turns out to be difficult
to proceed from the 1D case, where the sum over paths is universal
({\it i.e.} depends only on the dimension of the space-time), and 
the world-line action is known for a wide class of potentials and 
external fields, to the 2D case under study. 
In the next Section, we shall discuss
various field-theoretical models, where the string effective 
action and/or the measure in the sum over world sheets are either already 
postulated {\it a priori} or can be derived.

\section{Field-theoretical models and their string representations}
As a natural origin for the QCD string effective action serves
the Stochastic Vacuum Model (SVM) of the QCD vacuum~\cite{3} 
(for a review see {\it e.g.}~\cite{2}). Within this model,
the effective action reads

\begin{equation}
\label{1}
{\cal S}[\Sigma(C)]=2\int\limits_{\Sigma(C)}^{}d\sigma_{\mu\nu}(x)
\int\limits_{\Sigma(C)}^{}d\sigma_{\mu\nu}(x')
D\left(\frac{(x-x')^2}{T_g^2}\right).
\end{equation}
Here, $D$ is one of the two coefficient functions, which parametrize
the bilocal gauge-invariant correlator of the field strengths, and 
$T_g$ is the so-called correlation length of the QCD vacuum, {\it i.e.}
the distance at which the nonperturbative part of the $D$-function 
decreases as ${\rm e}^{-|x|/T_g}$. According to the existing lattice data,
$T_g\simeq 0.13{\,}{\rm fm}$ for the $SU(2)$-case~\cite{4}, and 
$T_g\simeq 0.22{\,}{\rm fm}$ for the $SU(3)$-case~\cite{5}
(see also Ref.~\cite{6} for related investigations and Ref.~\cite{7} 
for reviews). Clearly, Eq.~(\ref{1}) means that the function $D$ plays the 
r\^ole of the propagator of a nonperturbative gluon, which propagates 
between the points $x$ and $x'$ lying on the string world sheet 
$\Sigma(C)$.

The simplest Abelian-type theory possessing the property of 
confinement is the 3D compact QED~\cite{8}. In this theory, however, 
this phenomenon is
caused by stochastic magnetic fluxes penetrating through the 
contour $C$, which are generated by magnetic monopoles. Consequently, 
the string effective action in this case can be derived, rather than 
postulated. The dual (disorder) scalar field, describing the grand canonical
ensemble of monopoles
in this theory, acquires a nonvanishing (magnetic) 
mass due to the Debye screening in the Coulomb gas of monopoles.
Being proportional to the square root of the Boltzmann factor 
of a single monopole, this mass is essentially nonperturbative,
{\it i.e.} depends nonanalytically on the electric coupling constant.
It can be shown~\cite{add} that in the approximation of a 
dilute monopole gas, the string effective action has the form 
similar to Eq.~(\ref{1}) with the function $D$ replaced by the 
propagator of the massive dual boson.

Another Abelian-type theories where confinement takes place
are the so-called Abelian-projected theories~\cite{9}. There, 
within the so-called Abelian dominance hypothesis~\cite{10}, one
gets as an effective IR theory,
corresponding to the $SU(N_c)$-gluodynamics, the $[U(1)]^{N_c-1}$
magnetically gauge-invariant dual theory with monopoles. 
Further investigations of such a theory depend on the particular
form of the average over the monopole ensemble. One of the possibilities
is to treat this ensemble as a Coulomb gas, which is
a good approximation in 3D. In particular, in the $SU(2)$-case one 
then arrives just at the compact QED, discussed above. Another 
possibility, more appropriate in 4D, 
is to demand the condensation of the monopole Cooper pairs,
which leads to the dual Abelian Higgs type theory.
In such a theory, 
confinement can be described as the dual Meissner effect~\cite{11},
{\it i.e.} it is due to the formation of the dual 
Nielsen-Olesen strings~\cite{12}. By introducing external particles,
electrically charged {\it w.r.t.} the Cartan subgroup, one can see
that the string effective action in this theory again has the 
form~(\ref{1}) with the function $D$ replaced by the propagator of a
dual vector boson. Contrary to the Debye mechanism realized in the 
monopole gas, the dual bosons acquire now their magnetic mass due to the 
Higgs mechanism.

It is further possible to perform an expansion of the nonlocal 
action~(\ref{1}) in powers of the derivatives {\it w.r.t.} the 
world-sheet coordinates. Clearly, in the SVM-case such an expansion
is equivalent to the $T_g$-one, whereas in the case of 
Abelian-projected theories this is just an expansion in the inverse 
powers of the magnetic mass of the dual boson. Then, as the two
leading terms of this gradient expansion 
one gets the Nambu-Goto term and the 
so-called rigidity term~\cite{13}, whose coupling constants in 
the SVM-case read~\cite{14} $\sigma=4T_g^2\int d^2zD\left(z^2\right)$ 
and $\frac{1}{\alpha_0}=-\frac{T_g^4}{4}\int d^2zz^2D\left(z^2\right)$,
respectively. The positive string tension $\sigma$ as well as the 
negative sign of the other coupling constant ensure the stability
of string configurations in the models under consideration
(see Ref.~\cite{15} for a detailed discussion).

Finally, it is worth remarking on the measure of the summation 
over string world sheets in the above-mentioned theories. In the SVM, 
it is yet unknown as well as in the QCD itself and presumably 
defined by the perturbative fluctuations of the vacuum, mentioned 
in the Introduction. In 3D compact QED, 
as it has been argued 
in Ref.~\cite{add}, the $\Sigma(C)$-independence 
of $\left<W(C)\right>$ is realized by the summation over branches
of a certain multivalued potential of the monopole densities.
The same is true~\cite{16} for the 
$SU(3)$-inspired 3D Abelian-projected theory, where monopoles
form a gas~\cite{17}. In the 4D dual Abelian Higgs type theories,
the summation over string world sheets appears from the integration 
over the multivalued part of the phase of the dual Higgs field,
which is only nonvanishing on these world sheets. In particular, 
the Jacobian, which appears when one passes from the 
integration over this multivalued part to the integration 
over world sheets, has been evaluated in Ref.~\cite{18}.

In the next Section, we shall concentrate ourselves on the 
applications of the above ideas to the evaluation of the 
bilocal field strength correlator, which plays the key r\^ole in the 
SVM. For concreteness, we shall evaluate it in the effective 
$SU(3)$-inspired Abelian-projected theories in 4D and 3D.

\section{The bilocal field strength correlator in the 4D SU(3)-inspired
dual Abelian Higgs type theory}
In the London limit of infinitely heavy dual Higgs fields
({\it i.e.} infinitely thin dual strings), the 
Euclidean action of the model under study with an external
quark of the color $c$ reads~\cite{19}

\begin{equation}
\label{lsu3}
S_c=\int d^4x\left[\frac14\left({\bf F}_{\mu\nu}+{\bf F}^{(c)}_{\mu\nu}
\right)^2+\frac{\eta^2}{2}\sum\limits_{a=1}^{3}(\partial_\mu\theta_a-
2g_m{\bf e}_a{\bf B}_\mu)^2\right].
\end{equation}
Here, $g_m$ is the magnetic coupling constant, related to the QCD one 
as $g_m=\frac{4\pi}{g}$, ${\bf B}_\mu=\left(B_\mu^1,B_\mu^2\right)$ are the 
magnetic fields dual to the diagonal gluons 
${\bf A}_\mu=\left(A_\mu^3,A_\mu^8\right)$,
and $\eta$ is the {\it v.e.v.} of the dual Higgs fields. The phases 
$\theta_a$'s of the latter ones contain both the multivalued part,
describing dual Nielsen-Olesen 
strings, and the single-valued one, describing perturbative
fluctuations around them. Since monopole charges are distributed over the 
lattice defined by the root vectors ${\bf e}_a$'s of $SU(3)$,
${\bf e}_1=(1,0)$, ${\bf e}_2=\left(-\frac12,-\frac{\sqrt{3}}{2}\right)$,
${\bf e}_3=\left(-\frac12,\frac{\sqrt{3}}{2}\right)$, 
the dual Higgs fields are not independent of each other, but 
their phases are rather subject to the constraint 
$\sum\limits_{a=1}^{3}\theta_a=0$. The field of an external
quark of the color $c=R,B,G$ (red, blue, green, respectively) is 
represented in Eq.~(\ref{lsu3}) by the field strength tensor
${\bf F}_{\mu\nu}^{(c)}$, which obeys the equation $\partial_\mu
\tilde {\bf F}_{\mu\nu}^{(c)}={\bf Q}^{(c)}j_\nu^{\rm e}$. 
Here, $j_\nu^{\rm e}(x)\equiv g\oint\limits_{C}^{}dx_\nu(\tau)
\delta(x-x(\tau))$, and ${\bf Q}^{(c)}$'s are the weights of the 
representation ${\bf 3}$ of ${}^{*}SU(3)$, {\it i.e.} the charges 
of quarks {\it w.r.t.} the Cartan subgroup $[U(1)]^2$, which read
${\bf Q}^{(R)}=\left(\frac12,\frac{1}{2\sqrt{3}}\right)$, 
${\bf Q}^{(B)}=\left(-\frac12,\frac{1}{2\sqrt{3}}\right)$, 
${\bf Q}^{(G)}=\left(0,-\frac{1}{\sqrt{3}}\right)$.

The string representation of the action~(\ref{lsu3}) 
has been derived in Refs.~\cite{20},~\cite{ma} (see also Ref.~\cite{21} 
for the generalization to the case when the $\Theta$-term is included),  
and the resulting action reads

$$
S_c=\pi^2\int d^4x \int d^4yD_m^{(4)}(x-y)\left[\eta^2\bar
\Sigma_{\mu\nu}^a(x)\bar\Sigma_{\mu\nu}^a(y)+\frac{1}{6\pi^2}
j_\mu^{\rm e}(x)j_\mu^{\rm e}(y)\right].
$$
Here, $m=\sqrt{6}g_m\eta$ is the mass of the dual vector bosons,
which they acquire due to the Higgs mechanism, and $D_m^{(4)}(x)\equiv
\frac{m}{4\pi^2|x|}K_1(m|x|)$ is the respective propagator, where from now on
$K_\nu$'s stand for the modified Bessel functions. 
We have also introduced the 
following linear combinations of the vorticity tensor currents: 
$\bar\Sigma_{\mu\nu}^a\equiv\Sigma_{\mu\nu}^a-
2s_a^{(c)}\Sigma_{\mu\nu}^{\rm e}$. Here, $\Sigma_{\mu\nu}^a(x)=
\int\limits_{\Sigma_a}^{}d\sigma_{\mu\nu}(x_a(\xi))\delta(x-x_a(\xi))$
is the vorticity tensor current defined on the closed string world sheet
$\Sigma_a$, and $\Sigma_{\mu\nu}^{\rm e}$ is the analogous expression defined 
on an arbitrary open string world sheet $\Sigma^{\rm e}$ 
bounded by the contour $C$.
The numbers $s_a^{(c)}$'s read 
$s_3^{(R)}=s_2^{(B)}=
s_1^{(G)}=0$, $s_1^{(R)}=s_3^{(B)}=s_2^{(G)}=
-s_2^{(R)}=-s_1^{(B)}=-s_3^{(G)}=1$ and obey the relation
${\bf Q}^{(c)}=\frac13{\bf e}_as_a^{(c)}$.  
Note also that owing to the constraint imposed on $\theta_a$'s, 
$\Sigma_{\mu\nu}^a$'s also obey the constraint 
$\sum\limits_{a=1}^{3}\Sigma_{\mu\nu}^a=0$, which should be imposed 
by the introduction of the respective $\delta$-function into the 
partition function.

Following the SVM,  
let us further parametrize the bilocal correlator of 
electric field strengths ${\bf f}_{\mu\nu}=\partial_\mu{\bf A}_\nu-
\partial_\nu{\bf A}_\mu$ as

$$\left<f_{\mu\nu}^i(x)f_{\lambda\rho}^j(0)
\right>_{{\bf A}_\mu, {\bf j}_\mu^{\rm m}}=\delta^{ij}\Biggl\{
\Biggl(\delta_{\mu\lambda}\delta_{\nu\rho}-\delta_{\mu\rho}
\delta_{\nu\lambda}\Biggr)\hat D\left(x^2\right)+$$

\begin{equation}
\label{co}
+\frac12\Biggl[\partial_\mu
\Biggl(x_\lambda\delta_{\nu\rho}-x_\rho\delta_{\nu\lambda}\Biggr)
+\partial_\nu\Biggl(x_\rho\delta_{\mu\lambda}-x_\lambda\delta_{\mu\rho}
\Biggr)\Biggr]\hat D_1\left(x^2\right)\Biggr\}. 
\end{equation}
Here, $\left<\ldots\right>_{{\bf A}_\mu}$ is the standard average over 
free gluons, and 
$\left<\ldots\right>_{{\bf j}_\mu^{\rm m}}$ is a certain average
over Abelian-projected monopoles, which are coupled 
to the field ${\bf B}_\mu$ dual to 
${\bf A}_\mu$. Referring the reader to 
Ref.~\cite{22} for an exact form of the latter average, note
here only that it describes the condensation
of monopole Cooper pairs. By virtue of the parametrization~(\ref{co}),
one has 

$$
-\ln\frac{\int {\cal D}x_\mu^a\delta\left(
\sum\limits_{a=1}^{3}\Sigma_{\mu\nu}^a\right)
{\rm e}^{-S_c}}
{\int {\cal D}x_\mu^a\delta\left(
\sum\limits_{a=1}^{3}\Sigma_{\mu\nu}^a\right)
{\rm e}^{-S_c[C=0]}}=\frac{1}{24}\int d^4x \int d^4y\times
$$

\begin{equation}
\label{par}
\times\Biggl[
2g^2\Sigma_{\mu\nu}^{\rm e}(x)\Sigma_{\mu\nu}^{\rm e}(y)
\hat D\left((x-y)^2\right)+
j_\mu^{\rm e}(x)j_\mu^{\rm e}(y)\int\limits_{(x-y)^2}^{+\infty}
d\lambda\hat D_1(\lambda)
\Biggr]. 
\end{equation}

Let us next specify the average over closed strings 
on the L.H.S. of Eq.~(\ref{par}).
It is known (see {\it e.g.}~\cite{23}) that in the case of zero 
temperature under study such strings 
form virtual bound states consisting of a string and an antistring
({\it i.e.} two strings with opposite winding numbers), which are 
called vortex loops. The typical sizes
of these objects are much smaller than the distances between them,
which enables one to treat their grand canonical ensemble in the 
dilute gas approximation. The effective disorder field theory 
describing this ensemble has been constructed in Ref.~\cite{24}, and 
its action is given by the following formula:

$$S=\int d^4x\left\{
\frac{1}{12\eta^2}
\left(H_{\mu\nu\lambda}^a\right)^2+\frac32 
g_m^2\left(h_{\mu\nu}^a\right)^2-\right.$$

\begin{equation}
\label{disord}
\left.
-2\zeta\left[\cos\left(\frac{\pi}{\Lambda^2\sqrt{2}}
\left|\sqrt{3}
h_{\mu\nu}^1+h_{\mu\nu}^2\right|\right)+ 
\cos\left(\frac{\pi}{\Lambda^2\sqrt{2}}\left|\sqrt{3}
h_{\mu\nu}^1-h_{\mu\nu}^2\right|\right)\right]\right\}.
\end{equation}
Here, $H_{\mu\nu\lambda}^a\equiv\partial_\mu h_{\nu\lambda}^a+
\partial_\lambda h_{\mu\nu}^a+\partial_\nu h_{\lambda\mu}^a$
is the field strength tensor of the Kalb-Ramond field~\cite{kr}
$h_{\mu\nu}^a$. Next,  
$\zeta\propto{\rm e}^{-S_0}$ 
is a Boltzmann factor of a single vortex loop
with $S_0$ standing for its action, equal to the string tension of 
a loop times its area. We have also introduced an UV momentum cutoff
$\Lambda\equiv\sqrt{\frac{L}{a^3}}$, where $a$ denotes a typical 
size of the loop, whereas $L$ stands for a typical distance between
loops in the gas, so that $a\ll L$.
The masses of the Kalb-Ramond fields 
following from Eq.~(\ref{disord}) read
$M_a^2=m^2+m_a^2$, 
where $m_1=\frac{2\pi\eta}{\Lambda^2}
\sqrt{3\zeta}$, $m_2=\frac{2\pi\eta}{\Lambda^2}\sqrt{\zeta}$
are the contributions brought about by the Debye screening 
of the dual vector bosons in the gas of electric vortex loops.

By virtue of the representation of the effective theory~(\ref{disord})
in terms of the integral over the densities of the vortex 
loops~\cite{24},\cite{22}, one can perform 
the average over 
the grand canonical ensemble of these objects
on the L.H.S. of Eq.~(\ref{par}).
Then, within 
the approximation that the typical sizes of vortex loops 
are completely negligible {\it w.r.t.} the area of $\Sigma^{\rm e}$, 
one has~\cite{20} (see also Ref.~\cite{25} for the respective
$SU(2)$-calculations):

\begin{equation}
\label{m}
\hat D=\frac{m^3}{4\pi^2}
\frac{K_1(m|x|)}{\left|x\right|},
\end{equation}
 
\begin{equation}
\label{m1}
\hat D_1=
\frac{m}{2\pi^2x^2}\Biggl[\frac{K_1(m|x|)}{\left|x\right|}
+\frac{m}{2}\Biggl(K_0(m|x|)+K_2(m|x|)\Biggr)\Biggr]. 
\end{equation}
In the IR limit, $\left|x\right|\gg\frac1m$, 
the asymptotic behaviors of the obtained 
coefficient functions read

\begin{equation}
\label{mas}
\hat D\longrightarrow\frac{m^4}{4\sqrt{2}\pi^{\frac32}}
\frac{{\rm e}^{-m\left|x\right|}}{\left(m\left|x\right|\right)^
{\frac32}},~~
\hat D_1\longrightarrow\frac{m^4}{2\sqrt{2}\pi^{\frac32}}
\frac{{\rm e}^{-m\left|x\right|}}{\left(m\left|x\right|\right)^
{\frac52}}.
\end{equation}
Comparing now Eqs.~(\ref{mas}) with the results of the lattice 
investigations~\cite{4},\cite{5},\cite{6},\cite{7}, one can easily see
that they match each other. In particular, it is clearly seen 
that for the bilocal correlator of gluonic field strengths of the 
same kind, the vacuum of the model~(\ref{lsu3}) exhibits a nontrivial
correlation length $T_g=\frac{1}{m}$.

Moreover, it is possible to improve on the results~(\ref{m}) and~(\ref{m1})
by keeping the size of the vortex loops finite. Then, taking into account
the contributions of the bilocal correlators of the vortex loops 
to the coefficient functions $\hat D$ and $\hat D_1$, one arrives
at the following modifications of these functions~\cite{22}:

\begin{equation}
\label{Dtot}
\hat D=\frac{m^2M_2}{4\pi^2}
\frac{K_1(M_2|x|)}{|x|},
\end{equation}

\begin{equation}
\label{D1tot}
\hat D_1=\frac{m_2^2}{\pi^2M_2^2|x|^4}+
\frac{m^2}{2\pi^2M_2x^2}\left[\frac{K_1(M_2|x|)}{|x|}+\frac{M_2}{2}
\left(K_0(M_2|x|)+K_2(M_2|x|)\right)\right].
\end{equation}
It is straightforward to see that when $m_2$ vanishes, {\it i.e.} one 
disregards the effect of screening of the dual vector bosons by the 
vortex loops, the old expressions~(\ref{m}) 
and~(\ref{m1}) for the 
functions $\hat D$ and $\hat D_1$ are recovered. We also see that 
since the screening enhances the mass of the dual vector bosons, 
the correlation length of the 
vacuum becomes also modified from $\frac{1}{m}$ to $\frac{1}{M_2}$.
The asymptotic behaviors~(\ref{mas}) change respectively to

$$\hat D\longrightarrow\frac{(mM_2)^2}{4\sqrt{2}
\pi^{\frac32}}\frac{{\rm e}^{-M_2|x|}}{(M_2|x|)^{\frac32}},~~
\hat D_1\longrightarrow
\frac{m_2^2}{\pi^2M_2^2|x|^4}+\frac{(mM_2)^2}{2\sqrt{2}\pi^{\frac32}}
\frac{{\rm e}^{-M_2|x|}}{(M_2|x|)^{\frac52}}.$$ 
It is remarkable that the screening leads to the appearance of the IR  
power-like part of the function $\hat D_1$. On the other hand,
one can check that the UV asymptotic behaviors of the functions 
$\hat D$ and $\hat D_1$ remain unaffected by the screening, as well
as the string tension.

\section{The bilocal field strength correlator in the 3D-gas of 
$SU(3)$ Abelian-projected monopoles}
The partition function of the 3D grand canonical ensemble of  
$SU(3)$ Abelian-projected monopoles reads~\cite{17}

\begin{equation}
\label{gr}
{\cal Z}=1+\sum\limits_{N=1}^{\infty}
\frac{\zeta^N}{N!}\left(\prod\limits_{n=1}^{N}
\int d^3z_n\sum\limits_{a_n=\pm 1,\pm 2,\pm 3}^{}\right)
\exp\left[-\frac{g_m^2}{4\pi}\sum\limits_{n<k}^{}
\frac{{\bf e}_{a_n}{\bf e}_{a_k}}{\left|{\bf z}_n-
{\bf z}_k\right|}\right].
\end{equation}
Here, $\zeta\propto\exp\left(-\frac{{\rm const}}{g^2}\right)$ is now the 
Boltzmann factor of a single monopole, and ${\bf e}_{-a}=-{\bf e}_a$.
The action of the respective disorder field theory has the 
form~\cite{16}

$$S=\int d^3x\left\{
\frac12(\nabla\chi_1)^2+\frac12(\nabla\chi_2)^2-\right.$$

\begin{equation}
\label{threegas}
\left.-2\zeta\left[\cos(g_m\chi_1)+\cos\left(\frac{g_m}{2}
(\chi_1+\sqrt{3}\chi_2)\right)
+\cos\left(\frac{g_m}{2}
(\chi_1-\sqrt{3}\chi_2)\right)\right]\right\}.
\end{equation}
According to this expression, the Debye masses 
of the dual fields $\chi_1$ and $\chi_2$ 
are equal to each other and read $m=g_m\sqrt{3\zeta}$.
 
To discuss the field strength correlators in the model~(\ref{threegas}),
it is useful to derive the string representation for the monopole 
part of the Wilson loop.
The result has the form of an integral over the monopole 
densities ${\bf j}$'s and reads~\cite{16}

$$
\left<W(C)\right>_{\rm m}=\frac{1}{3{\cal Z}}\sum\limits_{c=R,B,G}^{}
\int {\cal D}{\bf j}\times
$$

\begin{equation}
\label{ti}
\times\exp\Biggl\{-\Biggl[
\frac{2\pi}{g^2}\int d^3x\int d^3y{\bf j}({\bf x})\frac{1}{|{\bf x}-
{\bf y}|}{\bf j}({\bf y})
+V[{\bf j}]-
i\int d^3x{\bf j}{\bf Q}^{(c)}
\eta\Biggr]\Biggr\}.
\end{equation}
Here, 

$$V[{\bf j}]=\sum\limits_{n=-\infty}^{+\infty}
\sum\limits_{\alpha=1}^{3}\int d^3x\times$$

\begin{equation}
\label{potential}
\times\left\{j_\alpha\left[\ln\left(\frac{j_\alpha}{2\zeta}+
\sqrt{1+\left(\frac{j_\alpha}{2\zeta}\right)^2}\right)+2\pi in\right]-
2\zeta\sqrt{1+\left(\frac{j_\alpha}{2\zeta}\right)^2}\right\}
\end{equation}
is the effective multivalued potential 
of the monopole densities with 
$j_1\equiv\frac{1}{\sqrt{3}}\left(\frac{1}{\sqrt{3}}j^1+j^2\right)$,
$j_2\equiv-\frac23j^1$, 
$j_3\equiv\frac{1}{\sqrt{3}}\left(\frac{1}{\sqrt{3}}j^1-j^2\right)$.
In Eq.~(\ref{ti}), we have also denoted by
$\eta$ the solid angle, under which 
the surface $\Sigma(C)$ shows up to an observer located at the 
point ${\bf x}$, 

$$\eta\left[{\bf x},\Sigma(C)\right]\equiv
\frac12\varepsilon_{\mu\nu\lambda}
\frac{\partial}{\partial x_\mu}\int\limits_{\Sigma(C)}^{}
d\sigma_{\nu\lambda}({\bf y})\frac{1}{|{\bf x}-{\bf y}|}.
$$
Notice that similarly to the case of 
compact QED~\cite{add}, it is the summation over the branches of the 
potential~(\ref{potential}), which restores the $\Sigma(C)$-independence
of $\left<W(C)\right>_{\rm m}$. Furthermore, one can restrict oneself
to the real branch of this potential by performing the replacement 
$\Sigma(C)\to\Sigma_{\rm min}(C)\equiv\Sigma_{\rm min}$. Then, 
in the dilute monopole gas 
approximation, $|{\bf j}|\ll\zeta$, it is straightforward to carry out
the resulting Gaussian integration over the monopole densities.
Combining the so-obtained result with the contribution to the 
Wilson loop stemming from the free diagonal gluons,

$$\left<W(C)\right>_{\rm free}=\exp\left(-\frac{g^2}{24\pi}
\oint\limits_{C}^{}dx_\mu\oint\limits_{C}^{}dy_\mu\frac{1}{|{\bf x}-
{\bf y}|}\right),$$  
we obtain the following result for the full Wilson loop:

$$\left<W(C)\right>=\left<W(C)\right>_{\rm m}
\left<W(C)\right>_{\rm free}=$$

$$
=\exp\left\{-\left[
\pi\zeta\int
\limits_{\Sigma_{\rm min}}^{}d\sigma_{\mu\nu}({\bf x})
\int\limits_{\Sigma_{\rm min}}^{}d\sigma_{\mu\nu}({\bf y})
+\frac{g^2}{24\pi}\oint\limits_{C}^{}dx_\mu\oint\limits_{C}^{}dy_\mu
\right]
\frac{{\rm e}^{-m|{\bf x}-{\bf y}|}}{|{\bf x}-{\bf y}|}
\right\}.
$$

Parametrizing further the bilocal correlator of the field strengths 
according to Eq.~(\ref{co}) (where the average 
$\left<\ldots\right>_{{\bf j}_\mu^{\rm m}}$ is now replaced by the 
average over the dilute 3D gas of monopoles)
with the redefinitions $\hat D\to{\cal D}$,
$\hat D_1\to{\cal D}_1$, we obtain:

$${\cal D}=12\pi\zeta\frac{{\rm e}^{-m|{\bf x}|}}{|{\bf x}|},~~
{\cal D}_1=\frac{24\pi\zeta}{(m|{\bf x}|)^2}\left(m+\frac{1}{|{\bf x}|}
\right){\rm e}^{-m|{\bf x}|}.$$
Similarly to the 4D-case considered in the previous Section, one 
can see a good correspondence of these results with those obtained
on the lattice for the real QCD in Refs.~\cite{4},\cite{5},\cite{6},\cite{7}.
In particular, the r\^ole of the correlation length of the vacuum
is played in the model under study by the inverse Debye mass, 
$\frac{1}{m}$, and at the distances $|{\bf x}|\gg\frac{1}{m}$,
${\cal D}\gg{\cal D}_1$ due to the preexponential behavior.

\section{Conclusions}
In this talk, we have addressed
the topic of a derivation of the bilocal field strength correlators
in various $SU(3)$-inspired 
Abelian-projected theories. Those included the 4D 
dual Abelian Higgs type theory and the 3D
theory, in which Abelian-projected monopoles were assumed to form a 
gas. In both cases, the derivation 
of the field strength correlators essentially employed the ideas
of the string representation of the respective theories. In particular,
in the first case the contribution of the bilocal correlators of 
vortex loops, formed by the closed dual strings with opposite 
winding numbers, to the correlator of the field strengths 
was accounted for. This led to the modification of the correlation 
length of the vacuum in the model under study. Namely, it changed 
from the classical expression, equal to the inverse 
mass of the dual vector bosons generated by the Higgs mechanism, 
to the inverse enhanced mass. The latter one 
took into account the effect of 
the Debye screening of these bosons in the gas of electric vortex loops.
Besides that, it turned out that the screening yielded also the 
IR power-like contribution to one of the two coefficient 
functions, which parametrized the bilocal correlator of the field 
strengths.

In the second case, the derivation of the field strength correlator
was essentially based on the string representation for the Wilson
loop in the 3D Coulomb gas of the $SU(3)$ Abelian-projected monopoles.
Similarly to the 4D dual Abelian Higgs type theory, 
in this case the results for the coefficient functions, which parametrize
the bilocal correlator of the field strengths, 
match the results 
of lattice measurements in QCD with a good accuracy. 
In both cases, only {\it w.r.t.} the correlators of the field strengths
built out of the gluon fields corresponding to the same generator of 
the Cartan algebra the vacua of the considered theories
exhibited a nontrivial correlation length.

\section*{Acknowledgments}
The author is grateful to Prof. A. Di Giacomo for useful discussions
and INFN for the financial support. He also acknowledges the organizers
of the fourth 
workshop ``Continuous Advances in QCD'' (Univ. Minnesota, May 2000)
for an opportunity to present these results in a very stimulating 
atmosphere.

\end{document}